\documentclass[twocolumn,
 reprint,
 amsmath,amssymb,
 aps,
 floatfix,
superscriptaddress, accepted=2019-06-14]{quantumarticle}

\usepackage{graphicx}
\usepackage{dcolumn}
\usepackage{bm}
\usepackage[subpreambles=true]{standalone}
\usepackage{hyperref}

\usepackage{array}
\usepackage{float}
\usepackage{calc}
\usepackage{textcomp}
\usepackage{mathtools}
\usepackage{multirow}
\usepackage{amsmath}
\usepackage{amssymb}
\usepackage{graphicx}

\newcommand{\ip}[2]{\langle #1, #2 \rangle}

\usepackage{braket}
\usepackage{qcircuit}
\usepackage[section]{placeins}
\usepackage{flafter}

\allowdisplaybreaks

\begin{document}

\title{Variational Quantum Computation of Excited States}

\author{Oscar Higgott}\thanks{oscar.higgott@gmail.com}
\affiliation{
Riverlane, 3 Charles Babbage Road, Cambridge CB3 0GT
}
\affiliation{
Department of Physics and Astronomy, University College London, London, WC1E 6BT
}
\orcid{0000-0001-9880-5218}
\author{Daochen Wang}
\affiliation{
Riverlane, 3 Charles Babbage Road, Cambridge CB3 0GT
}
\affiliation{Joint Center for Quantum Information and Computer Science,
University of Maryland, College Park, MD 20742}
\author{Stephen Brierley}
\affiliation{
Riverlane, 3 Charles Babbage Road, Cambridge CB3 0GT
}

\begin{abstract}
The calculation of excited state energies of electronic structure Hamiltonians has many important
applications, such as the calculation of optical spectra and reaction rates. While low-depth quantum
algorithms, such as the variational quantum eigenvalue solver (VQE), have been used to determine
ground state energies, methods for calculating excited states currently involve the implementation
of high-depth controlled-unitaries or a large number of additional samples. Here we show how
overlap estimation can be used to deflate eigenstates once they are found, enabling the calculation
of excited state energies and their degeneracies. We propose an implementation that requires the
same number of qubits as VQE and at most twice the circuit depth. Our method is robust to
control errors, is compatible with error-mitigation strategies and can be implemented on near-term
quantum computers.
\end{abstract}

\maketitle

\section{Introduction}

Eigenvalue problems are ubiquitous in almost all fields of science and engineering. Google's PageRank algorithm alone has had a significant impact on modern society, and at its core solves an eigenvalue problem associated with a stochastic matrix describing the World Wide Web~\cite{page1999pagerank}. Another important example is Principal Component Analysis (PCA) \cite{pearson1901liii, hotelling1933analysis}, which has widespread applications in bioinformatics, neuroscience, image processing, and quantitative finance. 

The time-independent Schr\"{o}dinger equation provides yet another example of a fundamental eigenvalue problem. Its numerical solution enables properties of atoms, molecules and materials to be predicted, with far-reaching applications in materials design, drug discovery and fundamental science~\cite{szabo2012modern}. Characterisation of excited state energies of molecules is required to predict charge and energy transfer processes in photovoltaic materials, or to understand some chemical reactions, such as those that involve photodissociation. However, classical methods such as density functional theory are often unable to determine excited states, even for materials where ground state energy calculations are possible.

Quantum computers have the potential to solve these and other problems significantly faster than any known methods using classical computers~\cite{lloyd2014quantum,shor1999polynomial, grover1996fast,aspuru2005simulated}. However many quantum algorithms will require quantum error correction, limiting their usefulness in the near future~\cite{preskill2018quantum}. Here we study hybrid quantum-classical algorithms, which dramatically reduce the required gate depth to run and somewhat mitigate errors, by closely integrating classical and quantum subroutines~\cite{mcclean2016theory,farhi2014quantum,wang2018generalised, johnson2017qvector,moll2017quantum, benedetti2019generative, mcardle2018variational, endo2018discovering}. 

The variational quantum eigensolver (VQE), introduced in Ref.~\cite{peruzzo2014variational}, is the first algorithm designed to find the lowest eigenvalue of a Hamiltonian on a near-term, non-fault-tolerant quantum computer. VQE is based on the variational principle and utilises the fact that quantum computers can store quantum states using exponentially fewer resources than required classically. VQE uses parameterised quantum circuits to prepare trial wavefunctions and compute their energy, and a classical computer to find the parameters minimising this energy. The low circuit depth of VQE has led to the hope that it may enable near-term quantum-enhanced computation.

Since its introduction, modifications have been suggested to enable VQE to find excited state energies: e.g.~a folded spectrum method~\cite{peruzzo2014variational} which requires finding the expectation of the squared Hamiltonian with quadratically more terms, or symmetry-based methods which are non-systematic~\cite{mcclean2016theory}. Such suggestions have been more recently superseded by two proposals:  a method that minimises the von Neumann entropy~\cite{santagati2018witnessing} and the quantum subspace expansion method~\cite{mcclean2017hybrid, colless2018computation}. However, the von Neumann entropy method (``WAVES'') requires a large number of high-depth controlled-unitaries, and the quantum subspace expansion method requires a large number of additional samples compared to VQE and introduces a new approximation.
 
Our algorithm extends VQE to systematically find excited states at almost no extra cost. We achieve this by adding ``overlap'' terms onto the optimisation function in order to exploit the fact that Hermitian matrices admit a complete set of orthogonal eigenvectors. Exploiting further the fact that VQE retains the classical parameters of ansatz states that enable their re-preparation, low-depth quantum circuits can then be readily used to calculate these overlap terms.

\section{Variational quantum deflation algorithm}\label{sec:vqda}

\begin{figure}
\centering 
  \includegraphics[width=0.9\columnwidth]{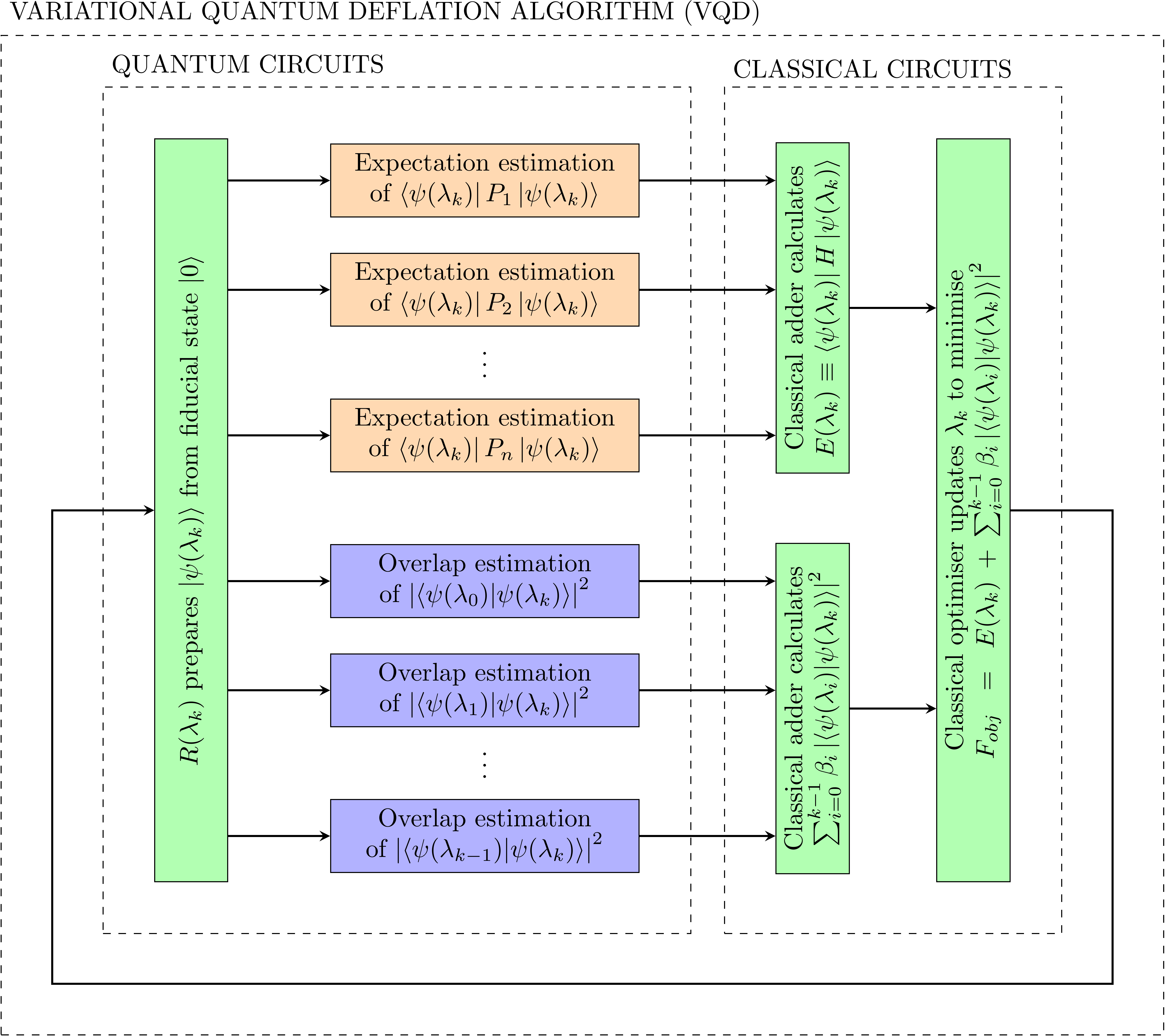}
  \caption{A schematic of our variational quantum deflation method for finding the $k$-th excited state of a Hamiltonian $H$.}
  \label{fig:vqd}
\end{figure}

In VQE, the real parameters ${\lambda}$ for the ansatz state $\ket{\psi({\lambda})}$ are classically optimised with respect to the expectation value:
\begin{equation}\label{eq:vqe}
E({\lambda})\coloneqq\bra{\psi({\lambda})}H\ket{\psi({\lambda})}=\sum_{j}c_{j}\bra{\psi({\lambda})}P_{j}\ket{\psi({\lambda})},
\end{equation}
 of the Hamiltonian $H=\sum c_{j}P_{j}$, computed using a low-depth quantum circuit. As a result of the variational principle, finding the global minimum of $E({\lambda})$ is equivalent to finding the ground state energy of $H$. VQE has been implemented on many experimental platforms, and has been shown to be more resilient to control errors than the quantum phase estimation algorithm~\cite{o2016scalable}. 
 
 Our method extends VQE to calculate the $k$-th excited state by instead optimising the parameters ${\lambda_k}$ for the ansatz state $\ket{\psi({\lambda_k})}$ such that the cost function:
 \begin{equation}\label{fobj}
F({\lambda_k})\coloneqq \bra{\psi({\lambda_k})}H\ket{\psi({\lambda_k})} + \sum_{i=0}^{k-1} \beta_i \left|\braket{\psi({\lambda_k})|\psi({\lambda_i})}\right|^{2},
\end{equation}
is minimised. This can be seen as minimising $E({\lambda_k})$ subject to the constraint that $\ket{\psi({\lambda}_k)}$ is orthogonal to the states $\ket{\psi({\lambda}_0)}, ..., \ket{\psi({\lambda}_{k-1})}$. In the next section, we show how choosing sufficiently large $\beta_0, ..., \beta_{k-1}$ means the minimum of $F({\lambda_k})$ is guaranteed to be the energy of the $k$-th state, provided that the ansatz is sufficiently expressive.

While the first term in Eq.~(\ref{fobj}) is $E({\lambda_k})$, and can be computed using the same quantum circuits as used for VQE, the second term is a sum of overlaps of the ansatz state with states $0$ to $k-1$, and can be computed efficiently on a quantum computer using one of the methods given in Section \ref{sec:lowdepth}.

Note that evaluating Eq.~(\ref{fobj}) requires knowledge of ${\lambda}_0, ..., {\lambda}_{k-1}$ and so an iterative procedure is required to calculate the $k$-th eigenvalue. First, ${\lambda}_0$ is calculated using VQE by minimising $E$ in Eq.~(\ref{eq:vqe}). Then, ${\lambda}_1$ is calculated by minimising $F$ in Eq.~(\ref{fobj}) for $k=1$, after which ${\lambda}_2$ can be determined using the same procedure with the known ${\lambda}_0$ and ${\lambda}_1$, and so on until ${\lambda}_k$ is determined.

A schematic of our variational quantum deflation (VQD) algorithm is shown in Fig.~\ref{fig:vqd}. An initial guess of ${\lambda}_k$ is used to generate a state preparation circuit $R({\lambda}_k)$ that prepares the state $\ket{\psi({\lambda_k})}$ when applied to the fiducial state $\ket{0}$. This circuit is used repeatedly to compute each of the expectation values $\bra{\psi({\lambda_k})}P_j\ket{\psi({\lambda_k})}$ (see Refs. \cite{peruzzo2014variational, wang2018generalised}) and overlap terms $\left|\braket{\psi({\lambda_k})|\psi({\lambda_i})}\right|^{2}$ for $i<k$. The overlap terms are computed using circuits described in Section \ref{sec:lowdepth} or Appendix~\ref{app:dest_swap} or by following the method in Ref.~\cite{KnillExpectationEstimation}.

A classical computer then uses the results of these quantum computations to calculate the objective function $F({\lambda_k})$ of Eq.~(\ref{fobj}) and update ${\lambda_k}$ using a classical optimiser. The new ${\lambda_k}$ is then used to prepare a new ansatz state on the quantum computer, and the whole process is repeated until some stopping criterion is reached.

As shown in Appendix~\ref{app:sampling_cost}, the total number of samples $M^{(k)}$ required to measure the VQD objective function to precision $\epsilon$ when finding the $k$\textsuperscript{th} excited state (assuming states $0\ldots k-1$ can be perfectly prepared) is bounded above by:
\begin{equation}
    M^{(k)}\leq \frac{1}{\epsilon^2}\left(\sum_{j=0}^{L-1}|c_j|+\frac{1}{2}\sum_{i=0}^{k-1}\beta_i\right)^2,
\end{equation}
compared to the VQE sampling cost of $M\leq \frac{1}{\epsilon^2}\left(\sum_{j=0}^{L-1}|c_j|\right)^2$. For well-chosen $\beta_i$, we expect this additional sampling cost relative to VQE to be very small, as explained in more detail in Appendix~\ref{app:sampling_cost}.

\section{Overlap weighting}

An equivalent viewpoint of our optimisation procedure is that we are finding the ground state of the effective Hamiltonian at stage $k$:
\begin{equation}\label{effH}
H_{k} \coloneqq H + \sum_{i=0}^{k-1}\beta_{i} \ket{i}\bra{i},
\end{equation}
where $\ket{i}$ is the (previously found) $i$-th eigenstate of $H$ with energy $E_{i}\coloneqq\bra{i}H\ket{i}$ \footnote{We assume these are true eigenstates with possibly non-distinct energies.}. It can be easily verified that for an arbitrary state $\Ket{\psi}\coloneqq\sum{a_i}\Ket{i}$:
$$\Bra{\psi}H_{k}\Ket{\psi} = \sum_{i=0}^{k-1} \lvert a_i \rvert ^2 (E_{i}+\beta_i) + \sum_{i=k}^{d-1} \lvert a_i \rvert ^2 E_{i},$$
where $d$ is the total number of eigenvectors of $H$.

Therefore, if the ansatz is sufficiently powerful, then to guarantee a minimum at $E_{k}$, it suffices to choose $\beta_{i} > E_{k} - E_{i} $. Since $\Delta \coloneqq E_{d-1}-E_{0} \geq E_{k}-E_{i}$,  it suffices to possess an accurate estimate of $\Delta$, e.g.~by using VQE to find $E_0$ and then $E_{d-1}$ (using the Hamiltonian $-H$ to find the latter). When we readily have a specification of $H=\sum c_{j}P_{j}$ as a linear combination of Pauli matrices, e.g.~when $H$ is the electronic structure Hamiltonian, then we have the upper bound $\Delta \leq 2 \lVert H \rVert \leq 2\sum \lvert c_{j}\rvert $. In this case, we can readily choose $\beta_i$ to guarantee the validity of our procedure. 

Choosing valid $\beta_i$ can also be self-correcting. For example, if we incorrectly chose $\beta_{i} = \gamma-E_i \leq E_{k} - E_{i}$ for all $i$, we will discover that we have set $\beta_i$ too small since we will eventually find a minimum at $F({\lambda_k})=\gamma$. However, by repeating the algorithm with a larger $\gamma$ until an energy strictly less than $\gamma$ is found (doubling $\gamma$ each time, say), we can pick a large enough $\gamma$ after $O(\log{(E_k-E_0)})$ runs of the algorithm.

\section{Low-depth implementations}\label{sec:lowdepth}

A low-depth method for overlap estimation, proposed in Ref.~\cite{havlicek2019supervised}, can be seen by writing the overlap $|\braket{\psi({\lambda}_i)|\psi({\lambda}_k)}|^2$ as $|\bra{0}R({\lambda}_i)^\dagger R({\lambda}_k)\ket{0}|^2$. We can prepare the state $R({\lambda}_i)^\dagger R({\lambda}_k)\ket{0}$ using the trial state preparation circuit followed by the inverse of the preparation circuit for the $i$-th previously-computed state. The overlap is then estimated to precision $\epsilon$ by the fraction of all-zero bitstrings when measuring this state $O(1/\epsilon^2)$ times in the computational basis.

This method requires knowing the inverse of the preparation circuit for each previously-computed state, $R({\lambda}_i)^\dagger$. While this inverse is often known in theory by inverting gates in a decomposition of the original preparation circuit, device errors may mean that the implementation is inaccurate in practice. If we define ${\lambda}_i^*$ to be the optimal parameters originally found to prepare the $i$-th state $R({\lambda}_i^*)\ket{0}$ using VQD, then its inverse can be found by fixing ${\lambda}_i^*$ and varying the trial state parameters ${\lambda}_i$ such that the overlap $|\bra{0}R({\lambda}_i)^\dagger R({\lambda}_i^*)\ket{0}|^2$ is maximised. This technique enables VQD to retain the robustness to control errors that is characteristic of VQE~\cite{o2016scalable}.

This implementation of VQD requires the same number of qubits as VQE and around twice the circuit depth. In Appendix~\ref{app:dest_swap}, we describe an alternative method which uses the destructive SWAP test and requires almost the same circuit depth as VQE but twice the number of qubits. If a larger gate-depth is available, then $\alpha$-QPE~\cite{wang2018generalised} can be used to reduce the total runtime of overlap estimation from $O(\frac{1}{\epsilon^2})$ up to $O(\frac{1}{\epsilon})$.

\begin{figure}[H]
\centering
\includegraphics[width=0.98\columnwidth]{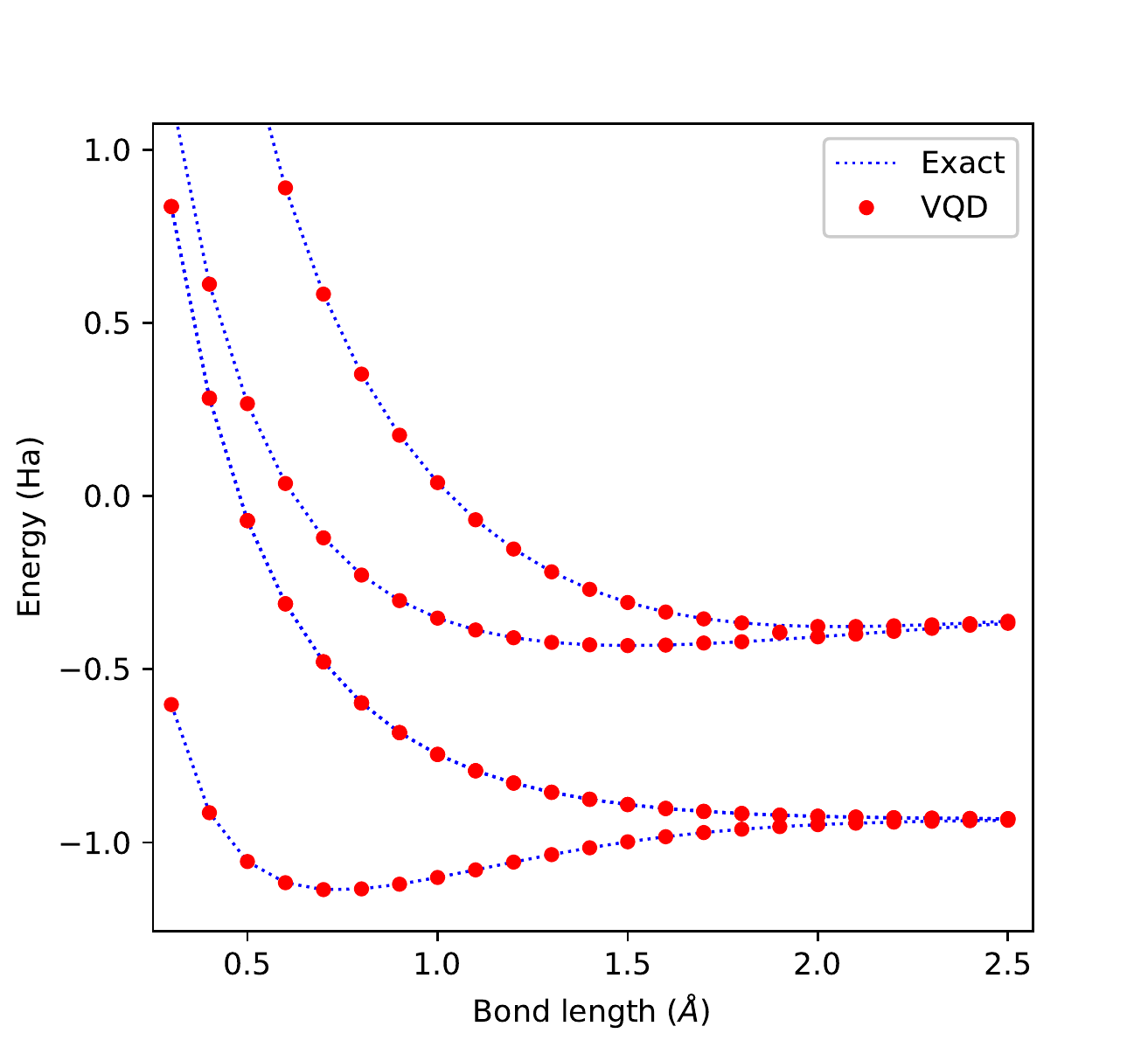}
\caption{All ground and excited state energy levels of H$_2$ in the STO-3G basis, calculated using exact diagonalisation (blue dotted line) and our variational quantum deflation (VQD) method (red filled circles) over a range of internuclear separations.}
\label{fig:numerics_h2}
\end{figure}

\section{Numerical simulation: H\textsubscript{2}}\label{sec:vqd_numerics}

We simulated VQD on H$_2$ in the STO-3G basis for a range of internuclear separations and compared it to exact diagonalisation, as shown in Fig.~\ref{fig:numerics_h2}. Using $\beta_i=3$ Ha for all $i$ and a generalised unitary coupled cluster singles and doubles (UCCGSD) ansatz, the median error of our method relative to exact diagonalisation is less than $4\times 10^{-6}$ Ha for all energy levels, significantly better than chemical accuracy of $1.6\times 10^{-3}$ Ha (the precision required to determine reaction rates to within an order of magnitude at room temperature using the Eyring equation~\cite{eyring1935activated}). Our method finds all 6 eigenstates systematically, including all those in the 3-dimensional degenerate subspace spanned by the 1st, 2nd and 3rd excited states. The ability to find degenerate states is another key advantage of our method; the folded spectrum and WAVES methods rely on the energies of states to differentiate between them and have no systematic way of determining the degeneracy of the eigenvalues. Further discussion of our simulation, including optimiser and ansatz used, can be found in Appendix~\ref{app:numerics}.

\section{Error accumulation}\label{sec:error_accumulation}

\begin{figure}
    \centering
    \includegraphics[width=\columnwidth]{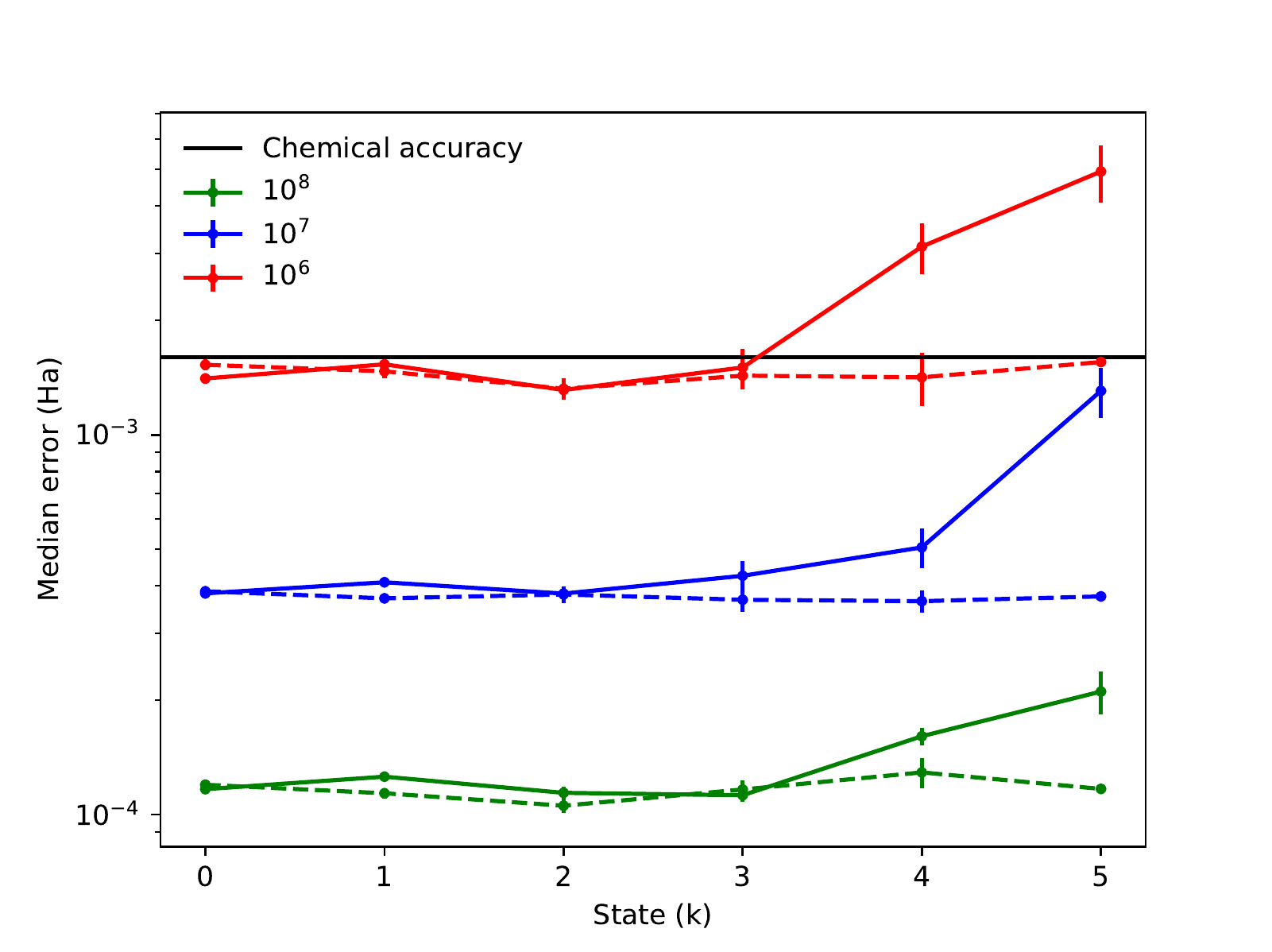}
    \caption{Median error using VQD to determine each energy level $k$ in the spectrum of $H_2$ at bond distance (0.7414 \r{A}) in the STO-3G basis. Red, blue and green lines show results using $10^6$, $10^7$ and $10^8$ samples (per Hamiltonian subterm and overlap term) respectively. For the solid lines, the standard VQD algorithm was used, whereas for dashed lines states $i<k$ in $H_k$ were computed exactly for comparison. Chemical accuracy ($1.6\times 10^{-3}$~Ha) is also shown for reference (black solid line). Error bars show 1$\sigma$ standard errors for the median estimates (calculated using bootstrap resampling~\cite{efron1992bootstrap}).}
    \label{fig:accumulation}
\end{figure}

In general, we cannot assume perfect state preparation for states $i<k$. Suppose a state $\ket{\tilde{\psi_0}}$ (with energy $\tilde{E_0}$) is prepared instead of the true ground state $\ket{\psi_0}$ such that $\left |\braket{\tilde{\psi_0}|\psi_0}\right |^2=1-\epsilon_0$, leading to an error in the ground state energy $\epsilon_0^\prime=\tilde{E_0}-E_0=O(2\epsilon_0 ||H||)$. If we now use this ground state estimate along with VQD to find the first excited state $\ket{\psi_1}$ using a new trial state $\ket{\tilde{\psi_1}}$, the lowest-energy state of the deflated Hamiltonian no longer corresponds to the exact excited state energy $E_1$. 

The inexact deflated Hamiltonian is now given by $\tilde{H_1}=H+\beta_0 \ket{\tilde{\psi_0}}\bra{\tilde{\psi_0}}$ and, to assess the accumulation of errors, we wish to find upper and lower bounds for $\min_{\tilde{\psi_1}}[\bra{\tilde{\psi_1}}\tilde{H_1}\ket{\tilde{\psi_1}}]$. 

In Appendix \ref{app:error_bounds} we show that, provided we set $\beta_0>\frac{E_1-E_0}{1-\epsilon_0}$, the ground state energy of the inexact deflated Hamiltonian is bounded by terms linear in $\epsilon_{0}$:
\begin{equation}
    E_1-O((E_1-E_0)\epsilon_0)\leq \min_\psi\bra{\psi}\tilde{H_1}\ket{\psi}\leq E_1 + \beta_0\epsilon_0.
\end{equation}

In reality we will not find the exact ground state of the deflated Hamiltonian $\tilde{H}_1$ and will incur an additional error $\epsilon_1$ as was the case for the ground state of the original Hamiltonian $H$. However, provided $\epsilon_1\approx \epsilon_0$, our total error $\epsilon_1^\prime=O(\epsilon_0\beta_0+2\epsilon_1 ||H||)$ in the energy is still linear in our original ground state error. An alternative analysis of error accumulation is provided by \citeauthor{lee2018generalized}~\cite{lee2018generalized}.

We analysed this accumulation of errors further through numerical simulations of VQD in the presence of sampling error, shown in Fig.~\ref{fig:accumulation}. We analysed three different sampling rates: $M=10^6$, $10^7$ and $10^8$ samples per Hamiltonian subterm and overlap term, running 225 simulations of VQD (with random initial parameters) for each of these three scenarios. Of these runs, $\sim 20\%$ of the simulations found the eigenstates in the incorrect order and were discarded for consistency in the analysis. The median errors for the remaining $\sim 180$ runs for each state $k$ are shown in Fig.~\ref{fig:accumulation}. For comparison, we also simulated 130 runs (dashed lines) using `exact` states $i<k$ in $H_k$ ($<10^{-7}$ energy error in each state $i<k$). For all three sampling rates, the median error in the first excited state is similar in magnitude to the error in the ground state, as expected from our analysis earlier in this section and in Appendix~\ref{app:error_bounds}. Furthermore, the median errors for all states $k<4$ are very similar (for a given $M$) to the errors when using an exact $H_k$, and are all below chemical accuracy, demonstrating that error accumulation is negligible for these states. For $k=4$ and $k=5$ the accumulated error is substantially higher than the error using an exact $H_k$, however, showing that VQD is most effective for low-lying states. Achieving chemical accuracy for $k=5$ requires $10^7$ samples, instead of $10^6$ for $0<k<4$.

One way to address this accumulation of errors within VQD to find higher excited states may be to use the alternative effective Hamiltonians discussed in Section~\ref{sec:effective_hamiltonians}. Another solution is to use a hybrid approach, using VQD instead of excitation operators in the WAVES protocol~\cite{santagati2018witnessing}. Here, VQD may provide a more effective method of approximating excited states than the excitation operators proposed in WAVES, whereas the von-Neumann entropy ``eigenstate witness'' used in WAVES does not have the same problem of error accumulation, and could help refine the energy estimate. Both of these alternative approaches require a larger gate depth than the version of VQD we have studied here, but may be a good approach to finding higher excited states in the era of fault-tolerant quantum computing.

\section{Choice of effective Hamiltonian}\label{sec:effective_hamiltonians}

The form of our effective Hamiltonian in Eq.~(\ref{effH}) is only one choice within the broad category of deflation methods. Such methods are typically employed to find eigenvalues and eigenvectors of positive semi-definite matrices, often covariance matrices in the context of PCA, starting from the largest eigenvalues.

To make direct use of deflation methods for positive semi-definite matrices, note that the  Hamiltonian $H' \coloneqq -H+E'$ for some $E'\geq E_{d-1}$, e.g. $E' = \lVert H \rVert $, is positive semi-definite. Under this transformation, we find that Hotelling's deflation corresponds to our method and would set $\beta_i=E'-E_{i}$ in Eq.~(\ref{effH}). 

Other deflation methods exist such as projection deflation or Schur complement deflation which are designed to address the problem of not obtaining true eigenstates at each stage. These two methods, in contrast to Hotelling's, ensure that the true ground state of the effective Hamiltonian at each stage does not overlap with the previously found eigenstate estimate irrespective of its accuracy. Empirically, these two methods have been found to perform better than Hotelling's in the context of PCA on some datasets \cite{NIPS2008_3575}. 

For example, in projection deflation, the effective Hamiltonian at stage $k$ is defined as:

\begin{equation}\label{effH_proj}
H_{k}  = A_{k}^{\dag} (H-E') A_{k},
\end{equation}
where:
\begin{equation}\label{A_k}
A_{k} \coloneqq \prod_{i=0}^{k-1}(1-\Ket{i}\Bra{i})\approx 1-\sum_{i=0}^{k-1}\Ket{i}\Bra{i},
\end{equation}
and the last approximation holds when the previously found eigenvectors $\Ket{i}$ are truly orthogonal.

With this approximation, writing $H$ again as a linear combination of Pauli matrices $P_{j}$, the value of $\Bra{\psi}H_k\Ket{\psi}$ for an ansatz $\Ket{\psi}$ is a linear combination of terms of forms: $\Bra{\psi} P_j \Ket{\psi}$, $\lvert \Braket{\psi|i} \rvert^2$ as in Hotelling's deflation, but now additionally $ \Braket{\psi|i}$, $\Bra{\psi}P_j\Ket{i}$ and $\Bra{i}P_j\Ket{l}$ (for $i, l < k$). Without this approximation, we also need to calculate $\Braket{i|l}$. The important point now is that all these additional terms can still be quantum computed, e.g.~following the method in Ref.~\cite{KnillExpectationEstimation}.

\section{Discussion}

We have introduced a new method--variational quantum deflation (VQD)--for calculating low-lying excited state energies of quantum systems using a quantum computer. Our method requires the same number of qubits as the variational quantum eigensolver (VQE) for ground state methods, at most twice the maximum circuit depth (for any given ansatz) and a negligible increase in the number of required measurements. By contrast, existing methods for quantum computing excited states require a large overhead in resources compared to ground state methods. 

While we used a Nelder-Mead optimiser and UCCSD ansatz in our simulation of molecular Hydrogen here, we note that many other optimisers and ansatz circuits can also be used for VQD. After the first version of this paper was released, interesting work by \citeauthor{endo2018discovering} compared the use of two different optimisation methods as applied to our protocol to calculate the spectrum of a Lithium Hydride molecule~\cite{endo2018discovering}. More recently, work by \citeauthor{lee2018generalized} showed that using a multi-determinental reference state or their k-UpCCGSD ansatz can improve the precision of finding the first excited state of N$_2$ using VQD~\cite{lee2018generalized}. Further work could include numerical analysis of different optimisers and ansatz circuits for use within VQD in the presence of noise, as well as the effectiveness of the alternative effective Hamiltonians presented in Section \ref{sec:effective_hamiltonians}. 

Given its low resource requirements and compatibility with error-mitigation techniques, we hope that VQD may enable the quantum-enhanced computation of excited state energies in the near-future.

\appendix

\section{Sampling cost}\label{app:sampling_cost}

In VQE, the variance $\epsilon^2$ in the energy expectation value $\braket{H}$ after using $M_j$ samples for the measurement of each subterm $\braket{P_j}$ in the Hamiltonian $H=\sum c_j P_j$ is bounded by~\cite{peruzzo2014variational, rubin2018application}:
\begin{align}\label{eq:vqesamples}
    \epsilon^2 &= \sum_{j=0}^{L-1}\frac{c_j^2\sigma_j^2}{M_j}\\
    &= \sum_{j=0}^{L-1}\frac{c_j^2(1-\braket{P_j}^2)}{M_j}\leq \sum_{j=0}^{L-1} \frac{c_j^2}{M_j}.
\end{align}
where $\sigma_j^2=\mathrm{Var}\left[\braket{P_j}\right]=\braket{P_j^2}-\braket{P_j}^2$ is the intrinsic variance of the projective measurement of $\braket{P_j}$. Using the method of Lagrange multipliers, \citeauthor{rubin2018application}~\cite{rubin2018application} showed that the optimal choice of $M_j$ to minimise the total number of samples $M=\sum_j M_j$ used to achieve precision $\epsilon$ is:
\begin{equation}
    M_j=\frac{1}{\epsilon^2}|c_j|\sigma_j\sum_{i=0}^{L-1}|c_i|\sigma_i,
\end{equation}
which leads to a total number of samples
\begin{align}
    M=\frac{1}{\epsilon^2}\left(\sum_{j=0}^{L-1}|c_j|\sigma_j\right)^2 \leq \frac{1}{\epsilon^2}\left(\sum_{j=0}^{L-1}|c_j|\right)^2.
\end{align}
Assuming perfect state preparation for states $i<k$ in VQD, we find that the variance of the energy expectation value $\braket{H_k}$ of the deflated Hamiltonian $H_k$ is instead given by:
\begin{align}\label{eq:vqdvariance}
    \epsilon^2 &=\sum_{j=0}^{L-1}\frac{c_j^2\sigma_j^2}{M_j^k}+\sum_{i=0}^{k-1}\frac{\beta_i^2\tilde{\sigma_i}^2}{\tilde{M}_{i}^k}\\
    &\leq \sum_{j=0}^{L-1}\frac{c_j^2}{M_j^k}+\sum_{i=0}^{k-1}\frac{\beta_i^2}{4 \tilde{M}_{i}^k}
\end{align}
where $M_j^k$ is the number of samples used for measuring $\braket{P_j}$, $\tilde{M}_{i}^k$ is the number of samples used to estimate the overlap of the ansatz with the $i$\textsuperscript{th} previously found state and $\tilde{\sigma_i}^2=\left |\braket{i|k}\right |^2(1-\left |\braket{i|k}\right |^2)$ is the intrinsic variance of this overlap measurement. From a straightforward extension of the Lagrange multiplier approach used by \citeauthor{rubin2018application}~\cite{rubin2018application}, we now find the optimal $M_j^k$ and $\tilde{M}_{i}^k$ for the deflated Hamiltonian $H_k$ to be:
\begin{align}
    M_j^k=\frac{1}{\epsilon^2}|c_j|\sigma_j\left(\sum_{l=0}^{L-1}|c_l|\sigma_l+\sum_{i=0}^{k-1}\beta_i\tilde{\sigma_i}\right),\\
    \tilde{M}_{i}^k=\frac{1}{\epsilon^2}\beta_i\tilde{\sigma_i}\left(\sum_{j=0}^{L-1}|c_j|\sigma_j+\sum_{l=0}^{k-1}\beta_l\tilde{\sigma_l}\right).
\end{align}
This leads to a total number of samples $M^{(k)}$ given by:
\begin{align}
    M^{(k)}&=\sum_{j=0}^{L-1}M_j^k+\sum_{i=0}^{k-1}\tilde{M}_{i}^k\\
    &=\frac{1}{\epsilon^2}\left(\sum_{j=0}^{L-1}|c_j|\sigma_j+\sum_{i=0}^{k-1}\beta_i\tilde{\sigma_i}\right)^2\\
    &\leq \frac{1}{\epsilon^2}\left(\sum_{j=0}^{L-1}|c_j|+\frac{1}{2}\sum_{i=0}^{k-1}\beta_i\right)^2.
\end{align}

From a comparison of $M^{(k)}$ with $M$ we expect the sampling overhead of VQD relative to VQE to be very small, since the sum of the $L=O(N^4)$ Hamiltonian coefficients will likely be far larger than the sum of well-chosen weights $\beta_i$ for low-lying excited states in practice. Furthermore, the variances $\tilde{\sigma_i}^2$ tend to zero at convergence. However, if we require precision $\epsilon$ throughout the optimisation rather than just at convergence, and if we choose $\beta_i=2\sum_{j=0}^{L-1}|c_j|$ since this always guarantees $\beta_i$ is large enough, then we find $M^{(k)}=(1+k)^2M$ (where we have used the upper bounds for both $M^{(k)}$ and $M$).

\section{Destructive SWAP test}\label{app:dest_swap}

\begin{figure}[H]
\centering
\includegraphics[width=0.9\columnwidth]{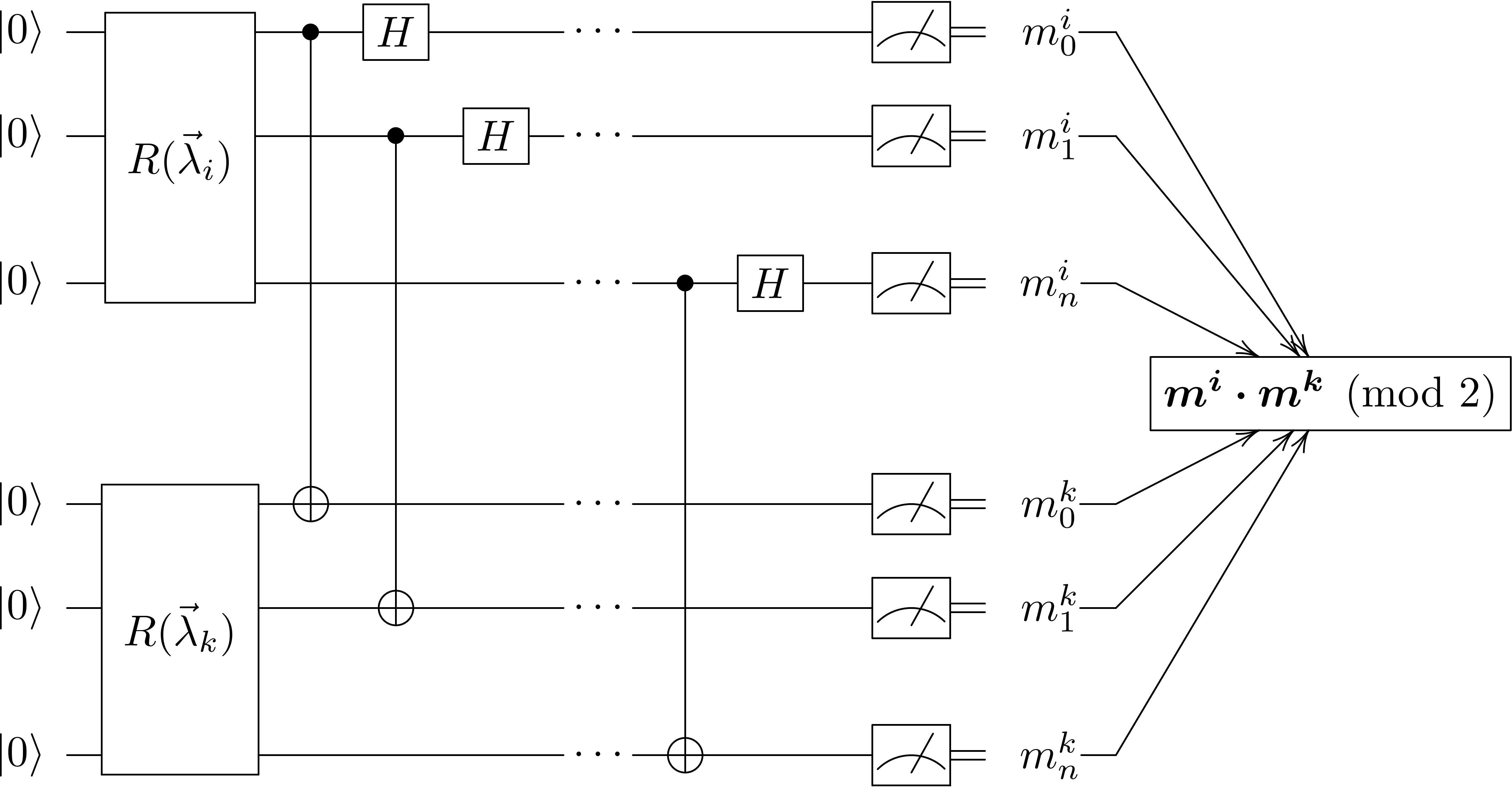}
\caption{The $N$-qubit generalisation of the destructive SWAP test as applied to two ansatz states $\ket{\psi({\lambda}_i)}$ and $\ket{\psi({\lambda}_k)}$, prepared using state preparation circuits $R({\lambda}_i)$ and $R({\lambda}_k)$ respectively.}
\label{fig:Nswpnqb}
\end{figure}

The SWAP test enables the overlap $\left|\braket{\phi|\psi}\right|^{2}$
of two states $\ket{\psi}$ and $\ket{\phi}$ to be determined to precision $\epsilon$ using $O(1/\epsilon^2)$ repeated measurements after applying a circuit to a quantum register
in the state $\ket{\psi}\otimes\ket{\phi}$. While the original SWAP test acting on two $N$-qubit states required an ancilla and a controlled-SWAP gate, leading to a $2N+1$-qubit circuit with depth $O(N)$, it was shown in Refs.~\cite{garcia2013swap,cincio2018learning} that the same outcome distribution can be attained more efficiently without an ancilla, using parallel Bell-basis measurements and classical logic. This so-called ``destructive SWAP test'' (shown in Fig.~\ref{fig:Nswpnqb}) requires just $2N$ qubits and depth $O(1)$, achieving significant savings compared to the original SWAP test. 

If the ansatz used can be implemented on a linear chain of qubits with nearest neighbour connectivity, e.g.~parameterised adiabatic state preparation using the fermionic SWAP network Trotter step~\cite{kivlichan2018quantum}, then the SWAP test to compare two ansatz states can be implemented on a $N\times2$ nearest-neighbour grid quantum computer architecture with a depth-one circuit that is subgraph isomorphic to the architecture (i.e.~no routing of quantum information required). This implementation makes the assumption that the same ansatz state can be prepared with the same parameters on two separate registers of qubits. If this cannot be assumed (e.g.~if qubit errors are inhomogeneous), then the SWAP test can be used to ``copy'' the state from the first register to the second register, by maximising the overlap of the two states, with the parameters of the state on the second register allowed to vary. This technique allows the SWAP test implementation of VQD to maintain robustness to control errors.

\section{Methods for numerical simulation}\label{app:numerics}

The standard UCCSD ansatz \cite{romero2017strategies} is defined relative to a reference state $\ket{\psi_0}$ by:
\begin{equation}\label{uccsd}
\ket{\psi} = e^{T-T^\dag}\ket{\psi_0},\nonumber
\end{equation}
where $T\coloneqq T_1+T_2$ with:
\begin{align}\label{uccsd}
T_1 &\coloneqq \sum_{\substack{i\in\text{occ} \\ l\in\text{vir}}}{t^l_i a_m^\dag a_i}, \nonumber\\
T_2 &\coloneqq \sum_{\substack{i,j\in\text{occ} \\ l,k\in\text{vir}}}{t^{lk}_{ij} a_l^\dag a_k^\dag a_i a_j},\nonumber
\end{align}
for some parameters $t^l_i$, $t^{lk}_{ij}\in\mathbb{R}$ and occ and vir are the sets of occupied and virtual orbitals of $\ket{\psi_0}$. 

We instead use a generalised unitary coupled cluster ansatz (UCCGSD) with $\Ket{\psi_{0}} = \Ket{\text{HF}}$ set to the Hartree-Fock state but with the cluster operator $T=T_1+T_2$ now using the definitions:
\begin{align}
T_1 &\coloneqq \sum_{pq}{t^q_p a_q^\dag a_p} \nonumber, \\
T_2 &\coloneqq \sum_{pqrs}{t^{rs}_{pq} a_r^\dag a_s^\dag a_p a_q},\nonumber
\end{align}
where $p,q,r,s$ can now index any orbital (irrespective of its occupation in the reference state). A variant of this ansatz was suggested by \citeauthor{mcclean2016theory}~\cite{mcclean2016theory} in the context of VQE, and UCCGSD has since been investigated numerically~\cite{lee2018generalized,wecker2015progress}. \citeauthor{lee2018generalized} found UCCGSD to perform significantly better than UCCSD in VQE for a number of small molecules~\cite{lee2018generalized}.

Since we are only interested in the parameterisation of $T-T^\dagger$, and fermionic operators obey the anti-commutation relations:
\begin{equation*}
\{a_j,a_k\}=0,\ \{a_j^\dag,a_k^\dag\}=0,\ \{a_j,a_k^\dag\}=\delta^k_j,
\end{equation*} 
it can be directly verified that there are only 6 and 3 independent parameters for $T_1$ (e.g. $t_0^1,\ t_0^2,\ t_0^3,\ t_1^2,\ t_1^3,\ t_2^3)$ and $T_2$ (e.g. $t_{01}^{23},\ t_{02}^{13},\ t_{03}^{12})$ respectively. 

The results in Fig.~\ref{fig:numerics_h2} were simulated using ProjectQ and FermiLib~\cite{steiger2018projectq, mcclean2017openfermion}. A tolerance of $10^{-2}$ was used with a Nelder-Mead optimiser (xatol=fatol=$10^{-2}$, as implemented in the scipy Python scientific library), and the best of two consecutive (randomly initialised) runs was used for each bond length and energy level. We note that other optimisers, such as LGO, have been shown to offer improved performance in VQE~\cite{mcclean2016theory}, and possible further work includes analysis of alternative optimisation strategies in the context VQD. 

While we initialised the UCCGSD parameters randomly in this work, choosing a good initial guess for the parameters can significantly reduce the number of iterations required for the optimiser to converge. For ground state VQE problems, second order M\o ller-Plesset perturbation theory (MP2) has previously been proposed as a UCC ansatz initialisation method~\cite{romero2017strategies}. Another method, for either ground or excited states, initialises a UCC ansatz with optimised expectation values that are classically estimated using a truncated BCH expansion~\cite{lee2018generalized}.

We also note that the overlap terms $|\braket{\psi({\lambda_k})|\psi({\lambda_i})}|^{2}$ in Eq.~(\ref{fobj}) of the state $k$ with a known state $i$ are similar to the overlap terms $|\braket{\psi({\lambda_s^t})|\psi({\lambda_i})}|^{2}$ of the same known state $i$ with another previously-computed state $s$ (where $i<s<k$) in the $t$-th iteration of the VQD optimisation procedure used to compute that state. It may therefore be advantageous to cache the outputs of these $|\braket{\psi({\lambda_s^t})|\psi({\lambda_i})}|^{2}$ terms, and use them to inform and improve the optimisation procedure for the $k$-th state, hopefully reducing the number of optimisation steps and quantum circuits required. 

\section{Bounds for error accumulation}\label{app:error_bounds}

In Section \ref{sec:error_accumulation} we stated that, to assess the accumulation of errors, we would like to find upper and lower bounds for the ground state energy $\min_{\psi}[\bra{\psi}\tilde{H_1}\ket{\psi}]$ of the inexact deflated Hamiltonian $\tilde{H_1}=H+\beta_0 \ket{\tilde{\psi_0}}\bra{\tilde{\psi_0}}$, where $\tilde{\psi_0}$ is the (inexact) estimate of the ground state found in the first iteration of VQD.

Using the same notation as in Section \ref{sec:error_accumulation}, and writing states in the eigenbasis of the Hamiltonian, $\ket{\tilde{\psi_0}}=\sum_{i=0}^{d-1}a_i\ket{\psi_i}$ and $\ket{\tilde{\psi_1}}=\sum_{i=0}^{d-1}b_i\ket{\psi_i}$, an $O(\beta_0\epsilon_0)$ upper bound is given straightforwardly by $\bra{\psi_1}\tilde{H_1}\ket{\psi_1}\leq E_1+\beta_0 \epsilon_0$. Writing $a_1^\rightarrow \coloneqq (a_1,\dots,a_{d-1})$ and $b_1^\rightarrow \coloneqq (b_1,\dots,b_{d-1})$ for compactness, we find the lower bound to be:

\begin{align}\label{eq:inaccurate_deflation_alt}
    &\bra{\tilde{\psi}_1}\tilde{H}_{1}\ket{\tilde{\psi}_1} \\
    &= |b_{0}|^2E_{0} + \sum_{i=1}^{d-1}|b_i|^2E_i + \beta_{0} |a_{0}^{*}b_{0} +  \ip{a_1^\rightarrow}{b_1^\rightarrow}|^2\\
    &= |b_{0}|^2(E_{0}+\beta_{0}|a_{0}|^2) + \sum_{i=1}^{d-1}|b_i|^2E_i \nonumber\\  & \quad +\beta_{0}(2\Re(a_{0}^{*}b_{0}\ip{a_1^\rightarrow}{b_1^\rightarrow}^{*})+|\ip{a_1^\rightarrow}{b_1^\rightarrow}|^2)\\
    &\geq |b_{0}|^2(E_{0}+\beta_{0}(1-\epsilon_{0})) - |b_{0}|(2\beta_{0}\sqrt{\epsilon_{0}})\nonumber\\
    & \quad + \min_{b_1^\rightarrow}\sum_{i=1}^{d-1}|b_i|^2E_i\\
    &\geq |b_{0}|^2(\beta_{0}(1-\epsilon_{0})-(E_1-E_0))\nonumber\\
    &\quad -|b_0|(2\beta_0\sqrt{\epsilon_0})+ E_1\\
    &\geq E_1-\epsilon_0\frac{\beta_0^2}{\beta_0(1-\epsilon_0)-(E_1-E_0)}.
\end{align}
where the first inequality is Cauchy-Schwarz, the second inequality follows from $\sum_{i=0}^{d-1}|b_i|^2=1$, and the third inequality follows by minimising a quadratic over $|b_0|$ (assuming $\beta_0>\frac{E_1-E_0}{1-\epsilon_0}$). From the Taylor series expansion in $\epsilon_0$ of the second inequality we find:
\begin{equation}
    \bra{\tilde{\psi}_1}\tilde{H}_{1}\ket{\tilde{\psi}_1}\geq E_1 - \frac{\beta_0(E_1-E_0)}{\beta_0-(E_1-E_0)}\epsilon_0 + O(\epsilon_0^2),
\end{equation}
from which it is clear that, for any fixed $\beta_0> \frac{E_1-E_0}{1-\epsilon_0}$, we have a lower bound of:
\begin{equation}
    \min_{\tilde{\psi}_1} \bra{\tilde{\psi}_1}\tilde{H}_{1}\ket{\tilde{\psi}_1}\geq E_1 - O((E_1-E_0)\epsilon_0).
\end{equation}

\section{Symmetry constraints}\label{app:symm}

It is often the case that the Hilbert space of the Hamiltonian being considered is larger than the Hilbert space relevant to the particular problem of interest. For example, consider the electronic structure Hamiltonian in second quantised form:
\begin{equation}
    H=\sum_{ij} h_{ij}a_i^\dagger a_j + \sum_{ijkl}h_{ijkl}a_i^\dagger a_j^\dagger a_k a_l,
\end{equation}
where $a_i^\dagger$ and $a_i$ are the fermionic creation and annihilation operators for an electron in the $i$-th spin orbital, and where the coefficients $h_{ij}$ and $h_{ijkl}$ denote the one- and two-electron integrals, respectively. After the Hamiltonian is transformed through the Jordan-Wigner or Bravyi-Kitaev transformation, converting creation and annihilation operators into qubit operators, the dimension of the Hilbert space remains $2^N$, where $N$ is the number of spin orbitals. However, if one is interested only in states with a particular symmetry, the dimension of the Hilbert space restricted only to these states can be much smaller, e.g.~$\binom{N}{\eta}=O(N^\eta)$ instead of $2^N$ if only $\eta$-electron states are of interest.

If we wish to apply VQD to find excited states of a molecular Hamiltonian with a particular symmetry, it is necessary that the ansatz state for a desired excited state, at the global minimum of Eq.~(\ref{fobj}), be contained entirely within the restricted Hilbert space of interest. One way of ensuring this is to use an ansatz that always conserves the correct symmetry. For example, the fermionic unitary coupled cluster ansatz we use in Section~\ref{sec:vqd_numerics} conserves the desired number of electrons ($\eta=2$) of neutral molecular Hydrogen for all input parameters.

Alternatively, penalty terms can be included in the objective function such that the ansatz state has the desired symmetry at the global minimum of the objective function~\cite{mcclean2016theory,ryabinkin2018constrained}. This leads to a modified objective function:
\begin{equation}\label{symm_fobj}
F_{C}({\lambda_k})\coloneqq F({\lambda_k}) + \sum_{i} \mu_i \big[\bra{\psi({\lambda_k})}\hat{C_i}\ket{\psi({\lambda_k})}-c_i\big]^2,
\end{equation}
where $\hat{C_i}$ are symmetry constraining operators (e.g.~$\hat{N_e}$, $\hat{S^2}$, $\hat{S_z}$) and $c_i$ are constants corresponding to their desired expectation values.

Clearly, by incorporating any of these techniques, we can find the excited states of a Hamiltonian constrained to any particular symmetry of interest.

\section{Error mitigation}\label{app:mitigation}

In Refs.~\cite{mcardle2018error, bonet2018low, ryabinkin2018constrained}, an error-mitigating post-processing procedure was introduced that uses the operators $\hat{C_i}$ (defined in Appendix~\ref{app:symm}) to detect and discard all measurements that violate a required symmetry for energy expectation circuits in VQE-type algorithms. This procedure can produce more accurate expectation values in the presence of bit-flip errors and some combinations of two-qubit errors. 

After the first version of this paper was released, Ref.~\cite{endo2018discovering} incorporated our VQD technique to calculate excited states using imaginary time evolution. The authors also proposed a method to detect symmetry-breaking errors when using the ancilla-based SWAP-test, by performing symmetry measurements on the ansatz registers while measuring the overlap with the ancilla. However, using the low-depth overlap estimation circuit given in Section \ref{sec:lowdepth}, we can detect and discard any error that does not commute with a symmetry operator $\hat{C_i}$ using classical post-processing alone, provided that $\hat{C_i}$ is diagonal in the computational basis and commutes with the ansatz. In the Jordan-Wigner and Bravyi-Kitaev encodings, the operators for the number of electrons $\hat{N_e}$, spin up electrons $\hat{N_\uparrow}$ and spin down electrons $\hat{N_\downarrow}$ are diagonal in the computational basis, allowing these quantities to be computed classically in post-processing for both encodings. For example, starting from $\Ket{0}$, the UCC ansatz is prepared by $R_{\text{UCC}}({\lambda}) = V({\lambda})R_{\text{HF}}$, where $R_{\text{HF}}$ prepares the Hartree-Fock state $\Ket{\text{HF}}$ and $V \coloneqq e^{T-T^\dagger}$ is the UCC operator. Now, rather than measuring the fraction of all-zero bitstrings after performing $R_{\text{UCC}}({\lambda}_i)^\dagger R_{\text{UCC}}({\lambda}_k)\ket{0}=R_{\text{HF}}^\dagger V({\lambda}_i)^\dagger V({\lambda}_k)R_{\text{HF}}\ket{0}$, we can instead measure the fraction of bitstrings corresponding to $\ket{\text{HF}}$ after performing $V({\lambda}_i)^\dagger V({\lambda}_k)R_{HF}\ket{0}$. Since $V$ conserves electron number, we know that all measured bitstrings that do not correspond to the correct electron number can be discarded as per the post-processing procedure. Therefore, this method for error-mitigated overlap estimation is more efficient than the ancilla-based method proposed in Ref.~\cite{endo2018discovering} if $\hat{C_i}$ is diagonal in the computational basis. We also note that the error-mitigation techniques proposed in Refs.~\cite{endo2017practical,temme2017error} can be readily applied to our algorithm.


\end{document}